\documentclass{ws-procs9x6}

\usepackage{graphicx}

\usepackage{amssymb}

\begin{document}


\title{Isospin in Reaction Dynamics. The Case of Dissipative Collisions at 
Fermi Energies}

\author{M. Di Toro}

\address{Laboratori Nazionali del Sud INFN, Phys.Astron. Dept. 
 Catania University\\
Via S. Sofia 62, \\ 
I-95123 Catania, Italy\\ 
E-mail: ditoro@lns.infn.it}

\maketitle

\abstracts
{A key question in the physics of
unstable nuclei is the knowledge of the $EOS$ for asymmetric nuclear
matter ($ANM$) away from normal conditions. 
We recall that the symmetry energy at low densities has important effects
on the neutron skin structure, while the knowledge in
high densities region is crucial for supernovae dynamics
and neutron star properties. The $only$ way to probe such region
of the isovector $EOS$ in terrestrial laboratories is through
very dissipative collisions of asymmetric (up to exotic) heavy ions
from low to relativistic energies. A general introduction to the 
topic is firstly presented. We pass then to a detailed discussion
on the $neck-fragmentation$ process as the main dissipative mechanism at
the Fermi energies and to the related isospin dynamics.
From Stochastic Mean Field simulations the isospin effects on all the 
phases of the reaction
dynamics are thoroughly analysed, from the fast nucleon emission to
the mid-rapidity fragment formation up to the dynamical fission of the
$spectator$ residues. Simulations have been performed with an increasing
stiffness of the symmetry term of the $EOS$.
 Some differences 
have been noticed, especially for the fragment charge asymmetry.
New isospin effects have been revealed from the correlation of fragment 
asymmetry with dynamical quantities at the freeze-out time.
A series of isospin sensitive observables to be further
measured are finally listed.}

\section{Introduction}

While we
are planning second and third generation facilities for
radioactive beams our basic knowledge of the symmetry
term of the $EOS$ is still extremely poor.
Effective interactions are obviously tuned to symmetry properties
around normal conditions and any extrapolation can be quite
dangerous. Microscopic approaches based on realistic $NN$
interactions, Brueckner or variational schemes, or on effective
field theories show a rather large variety of predictions.
While all the potential symmetry energies  obviously cross at normal 
density $\rho_0$, quite 
large differences are present for values, slopes and curvatures
in low density and particularly in high density
regions.
Moreover even at the relatively well known ``crossing point'' at normal
density the various effective forces are presenting controversial
predictions for the momentum dependence of the fields acting on 
the nucleons and consequently for the splitting of the neutron/proton
effective masses, of large interest for nuclear structure and dynamics.

In the recent years under the stimulating perspectives offered from
nuclear astrophysics and from the new Radioactive Ion Beam ($RIB$) facilities
a relevant activity has started in the field of the isospin degree of freedom
in heavy ion reactions, see \cite{BaoJMPE7,Isospin01,DitoroEPJA13}.
Here we quickly review the field trying to pin down the most interesting 
theory
questions and eventually the related key observables.

The slope and curvature of the density dependence of the symmetry term
around saturation is related
to physics properties of exotic nuclei, bulk densities, neutron distributions
and monopole frequencies. The momentum dependence of the interactions in the 
isovector channel can be analysed discussing the effects
on the energy-slope of the Lane Potential at normal density and
on the symmetry field seen by high momentum nucleons \cite{Erice98,BaranPR05}.

The study of symmetry properties at
low density is important for 
the Liquid-Gas Phase
Transition in Asymmetric Matter.
A dynamical approach of great relevance since
it leads to the space-time properties of the unstable normal modes
which give rise to fragments inside the extended spinodal boundary. 
Characteristic coupled dispersion relations must be solved.
We remind that the coupling of the isoscalar and isovector
collective response is well known for stable modes \cite{GR98,HamamotoPRC60}
in neutron rich exotic nuclei, 
with transition densities that show a mixed nature. We can extend the same
results to the unstable responses of interest in fragmentation reactions. 
The mixed nature of the unstable
density oscillations will naturally lead to the {\it isospin distillation}
effect, i.e. a different {\it concentration} ($N/Z$) between the liquid
and the gas phase \cite{MuellerPRC52,BaranPRL86,ColonnaPRL88,MargueronPRC67}.
We can expect that quantitatively this effect will be dependent
on properties of the symmetry energy in very dilute matter, well below the
saturation point.

The early emission of particles in heavy ion collisions is
relevant for understanding both the reaction dynamics and the mechanism of
particle production. The density 
behavior of the symmetry potential largely influences the value of the
pressure that is governing the early reaction dynamics
 \cite{DanielNPA685}. Therefore 
we can predict important symmetry effects
on transport properties ( fast particle emission, collective flows)
in the asymmetric $NM$ that will
be probed by Radioactive Beam collisions at Fermi and intermediate energies.
 Different 
parametrizations in the momentum dependence of the isovector channel
can be  tested, in particular the transport effect of
the sign of the $n/p$ effective mass splitting $m^*_n-m^*_p$ in
asymmetric matter at high baryon and isospin density 
\cite{RizzoNPA732,BaranPR05}.

The Fermi 
energy domain represents the transition
region between a dynamics mainly driven by the mean-field,  
 below $15-20~AMeV$, 
and the one where the nucleon-nucleon collisions play a central role, 
above $100~AMeV$. We can have then a very rich variety of dissipative
reaction mechanisms, with related different isospin dynamics.
One of the new distinctive features is the enhanced production of
Intermediate Mass Fragments $(IMF, 3 \le Z \le 20)$. We analyse then
with great detail the isovector channel effect on the onset of the 
fragmentation
mechanism and on the isotopic content of the produced fragments.
In the reaction simulations it is essential the use of a 
{\it Stochastic Transport Approach} since fluctuations, instabilities and 
dynamical branchings are very important in this energy range
\cite{ColonnaNPA642,ColonnaNPA742}.
With the centrality of the collision we can have different 
scenarios for
fragment production, from the growing instabilities of dilute matter in central
reactions to the cluster formation at the interface between low and normal 
density regions in semicentral collisions. We can then expect a large variety
of isospin effects, probing different regions of the symmetry energy
below saturation. 

The traditional approach to nuclear physics starts from non-relativistic
formalisms in which non-nucleonic degrees of freedom are integrated out
giving nucleon-nucleon potentials. Then nuclear matter is described as a 
collection of quantum nonrelativistic nucleons interacting through an 
istantaneous effective potential. Although this approach has had a great
success, a more appropriate set of degrees of freedom consists of strongly
interacting effective hadron fields, mesons and baryons. These variables
are the most efficient in a wide range of densities and temperatures and
they are the degrees of freedom actually observed in experiments, in
particular in Heavy Ion Collisions at intermediate energies. 
Moreover this framework appears in any case a fundamental ``Doorway Step''
towards a more microscopic understanding of the nuclear matter.
Relativistic contributions to the isospin physics for static properties 
and reaction dynamics deserve an appropriate discussion.

The $QHD$ (Quantum-Hadro-Dynamics) effective field model represents a 
very successful attempt to 
describe, in a fully consistent relativistic picture, equilibrium and 
dynamical properties of nuclear systems at the hadronic 
level \cite{WaleckaAP83,SerotIJMPE6}.
 The same isospin physics described in detail at the
non-relativistic level can be naturally reproduced. Moreover some new
genuine relativistic effects can be revealed, due to the covariant structure
of the effective interactions.
One of the main points of the discussion is the relevance of the coupling
to a scalar isovector channel, the effective $\delta[a_0(980)]$ meson, not
considered in the usual nuclear structure studies
\cite{LiuboPRC65,GrecoPRC67}. 
 Particular attention is
devoted to the expectations for the splitting of the nucleon effective
masses in asymmetric matter, with a detailed analysis of the
relationship between the Dirac masses of the effective field approach and
the Schr\"odinger masses of the non-relativistic models.
A relativistic Dirac-Lane potential is deduced, which shows a structure
very similar to the Lane potential of the non-relativistic optical
model, but now in terms of isovector self-energies and coupling constants
 \cite{BaranPR05}.

The Relativistic Mean Field ($RMF$) approximation is allowing
more physics transparent results, often even analytical. We must
always keep a close connection to the more microscopic 
Dirac-Brueckner-Hartree-Fock ($DBHF$) approaches, in their extension to
asymmetric matter, \cite{LeePRC57,JongPRC57,HofmannPRC64,DalenarNPA744}
It is
well known that correlations are naturally leading to a density
dependence of the coupling constants, see ref.\cite{JongPRC57} for the
$DBHF$ calculations
and ref.\cite{GrecoPRC64} just for the basic Fock correlations.
Within the $RMF$ model we can get a clear qualitative estimation of the 
contribution
of the various fields to the nuclear dynamics. The price to pay is that
when we try to get quantitative effects we are forced to use different
sets of couplings in different baryon density regions.
In this case we can use the $DBHF$ results as guidelines
 \cite{GrecoPLB562,GaitanosNPA732,GaitanosPLB595}.

Results for relativistic Heavy Ion Collisions ($HIC$) in the $AGeV$
beam energy region, for collective flows, charged
pion production and isospin stopping, are presented 
in refs. \cite{GrecoPLB562,GaitanosNPA732,GaitanosPLB595}.
We recall that intermediate energy $HIC$'s represent the only way
to probe in terrestrial laboratories the in-medium effective interactions
far from saturation, at high densities as well as at high momenta.
Within a relativistic
transport model it is shown that the
isovector-scalar $\delta$-meson, which affects the high density
behavior of the symmetry term and the nucleon effective mass splitting, 
influences the
isospin dynamics. The effect is largely enhanced by a relativistic
mechanism related to the covariant nature of the fields
contributing to the isovector channel. The possibility is emerging of
a direct measurement of the Lorentz structure of the effective
nuclear interaction in the isovector channel. 
Very sensitive quantities appear to be the elliptic flow, related to 
the time scale of the particle emissions, and the isospin transparency in 
central collisions.

\section{Neck Dynamics at the Fermi Energies: Theory Survey}

The possibility of observation of new effects, beyond the Deep-Inelastic
binary picture, in fragment formation for semicentral collisions with 
increasing energy was advanced on the basis of the reaction dynamics
studied with transport models \cite{BertschPLB141,ColonnaNPA541,SobotkaPRC50}.
The presence of a time matching between the instability growth in the dilute
overlap zone and the expansion-separation time scale was suggesting
the observation of mean field instabilities first at the level of
anomalous widths in the mass/charge/... distributions of Projectile-Like
or Target-Like ($PLF/TLF$) residues in binary events, then through a direct 
formation
of fragments in the $neck-region$, \cite{ColonnaNPA589,Erice98}.
It is clear that in the transport simulations stochastic terms should be
consistently built in the kinetic equations in order to have a correct
description of instability effects. Stochastic Mean Field approaches
have been introduced, that well reproduce the presently available data
and reveal a large predictive power,
\cite{DitNPA681,BaranNPA703,BaranNPA730,BaranPR05}.

In conclusion at the Fermi energies we expect an interplay between binary
and $neck-fragmentation$ events, where Intermediate Mass Fragments ($IMF$, 
in the range $3 \le Z \le 10$) are directly formed in the overlapping region, 
roughly at mid-rapidity in semicentral reactions. The competition 
between the two mechanisms is expected to be rather sensitive to the
nuclear equation of state, in particular to its compressibility  that will
influence the interaction time as well as the density oscillation in the 
$neck-region$. In the case of charge asymmetric colliding systems the 
poorly known stiffness of the symmetry term will also largely influence 
the reaction dynamics. An observable sensitive to the stiffness of the 
symmetry term can be just the relative yield of incomplete fusion vs. 
deep-inelastic vs neck-fragmentation events \cite{ColonnaPRC57}.
Moreover for n-rich systems in the asy-soft case we expect more interaction 
time available for charge equilibration. This means that even the binary 
events will show a sensitivity through a larger isospin diffusion. At variance,
in the asy-stiff case the two final fragments will keep more memory of the
initial conditions. 

Systematic transport studies of isospin effects in the neck dynamics have been
performed so far for collisions of Sn-Sn isotopes at 50AMeV
\cite{DitNPA681,BaranNPA703}, Sn-Ni isotopes at 35AMeV 
\cite{BaranNPA730,BaranPR05} and finally Fe-Fe and Ni-Ni, mass 58, at
30 and 47 AMeV \cite{LiontiarX05}.

\begin{figure}
\centering
\includegraphics*[scale=0.35]{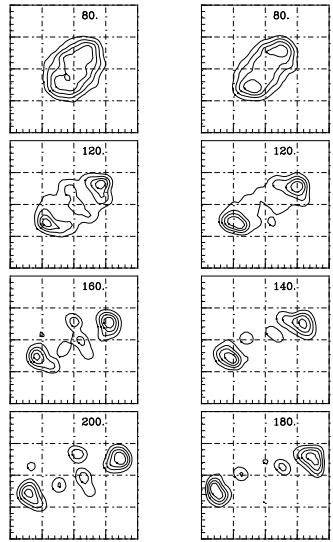}
\caption{$^{124}Sn+^{124}Sn$ collision at $50AMeV$: time evolution of the
nucleon density projected on the reaction plane.
Left column: $b=4fm$. Right column: $b=6fm$.}
\label{iso6}
\end{figure}

We show in Fig. \ref{iso6} the density contour plots
of a neck fragmetation event at $b=6fm$, for the
reaction $^{124}Sn+$$^{124}Sn$ at $50AMeV$, obtained
from the numerical calculations based on the stochastic transport
approach described before, \cite{BaranNPA703}.

\begin{figure}
\centering
\includegraphics*[scale=0.40]{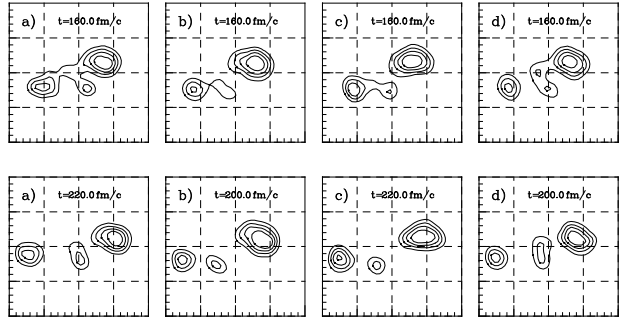}
\caption{Density contour plots for $^{124}Sn$+$^{64}Ni$ at $35 AMeV$ 
($b=6fm$). Early stage of fragment 
formation (top)
and same events close to the freeze-out (bottom) for four ternary cases a), b), c), d)
in neck fragmentation.}
\label{neck3}
\end{figure}

We clearly see neck-fragmentation events even at lower energies, now 
mostly ternary. For the system $^{124}Sn$+$^{64}Ni$ at $35 AMeV$ ($b=6fm$),
neck instabilities favour the appearance of $NIMF$'s,
after $150fm/c$, in a variety of places and ways as can be seen
by looking at Fig.\ref{neck3}, \cite{BaranNPA730}. For four events, we 
illustrate 
two characteristic stages, the early phase of the fragment formation process
and the configuration close to freeze-out.

\section{Neck Dynamics at the Fermi Energies: Experimental Survey}

It is now quite well established that the largest part of the reaction
cross section for dissipative collisions at Fermi energies goes
through the {\it Neck Fragmentation} channel, with $IMF$s directly
produced in the interacting zone in semiperipheral collisions on very short
time scales. We can expect different isospin effects for this new
fragment formation mechanism since cluster are formed still in a dilute
asymmetric matter but always in contact with the regions of the
Projectile-Like and Target-Like remnants almost at normal densities.

A first evidence of this new dissipative mechanism was suggested
at quite low energies, around $19AMeV$, in semicentral $^{100}Mo+^{100}Mo$,
$^{120}Sn+^{120}Sn$ reactions \cite{CasiniPRL71,StefaniniZFA351}.
A transition from binary, deep-inelastic, to ternary events was observed,
with a fragment formed dynamically 
that influences the fission-like decay of the primary 
projectile-like ($PLF$) and target-like ($TLF$) partners. 
Similar conclusions were reached in \cite{TokePRL75}.
Consistent with
the dynamical scenario was the anisotropic azimuthal distribution of IMF's.  
In fact the $IMF$ {\it alignement} with respect to the ($PLF^*$) velocity
direction has been one clear property of the ``neck-fragments'' first
noticed by Montoya et al.\cite{MontoyaPRL73} for 
$^{129}Xe + ^{63,65}Cu$ at $50AMeV$.

A rise and fall of the neck  mechanism for mid-rapidity fragments
with the centrality,  with a maximum for intermediate
impact parameters $b \simeq \frac{1}{2} b_{max}$, 
as observed in \cite{LukasikPRC55},
suggests the special
physical conditions required.
The size of the participant zone is of course important but it  
appears that a good time matching between the reaction and the 
neck instabilities time-scales is also needed, as suggested in
refs. \cite{BertschPLB141,ColonnaNPA589}. 
In fact a simultaneous presence, in noncentral collisions,
of different $IMF$ production mechanisms
at midrapidity was inferred in several experiments 
\cite{LefortNPA662,PlagnolPRC61,LarochellePRC62,GingrasPRC65,DavinPRC65,BocageNPA676,ColinPRC67,PaganoNPA734}. 

We can immediately predict an
important isospin dependence of the neck dynamics, from the presence of 
large density
gradients and from the possibility of selecting various time-scales
for the fragment formation. 
The first evidences of isospin effects in neck fragmentation
were suggested by Dempsey et al. in \cite{DempseyPRC54} from
semiperipheral collisions of the systems  $^{124,136}Xe +$ $^{112,124}Sn$ 
at $55AMeV$, where
 correlations between the average number of $IMF$'s,
$N_{IMF}$, and  neutron and charged particle multiplicities 
were measured. 
The variation of the relative yields of
$^{6}He/$$^{3,4}He$, $^{6}He/Li$
with  $v_{par}$ for several $Z_{PLF}$ gates
shows that the fragments produced in the midvelocity
region are more neutron rich than are the fragments
emitted by the $PLF$. 
Enhanced $triton$ production an midrapidity was considered
in  ref.\cite{LukasikPRC55}, and more recently in \cite{PoggiNPA685},
 as an indication of a neutron neck enrichement.

P.M. Milazzo et al. \cite{MilazzoPLB509,MilazzoNPA703} analyzed the
the $IMF$ parallel velocity distribution for $^{58}Ni$$+ ^{58}Ni$
semiperipheral collision at $30AMeV$. The two-bumps structure 
for $IMF$s with $5 \le Z \le 12$,
located around the center of mass velocity and close
to the quasiprojectile ($PLF^*$) source respectively,
was explained assuming the simultaneous presence of two 
production mechanisms:
the statistical disassembly of an equilibrated $PLF^*$ and 
the dynamical fragmentation of the participant region.
The average elemental event multiplicity $N(Z)$
exhibits a different trend for the two processes: more n-rich
in the mid-rapidity source..
This experiment has a particular importance since 
isospin effects were clearly observed, {\it in spite of the very low
inital asymmetry}.
The same reactions have been recently studied at the Cyclotron Inst.
of Texas $A\&M$ at various beam energies with measurements of the correlations
of fragment charge/mass vs. dynamical observables (emission angles and 
velocities) \cite{ShettyPRC68R,ShettyPRC70}.

\subsection*{Isospin Diffusion}

The isospin equilibration appears of large interest
even for more peripheral collisions, where we have 
shorter interaction times, less overlap and a competition
between binary and neck-fragmentation processes.  The low density
neck formation and the  preequilibrium emission are adding 
essential differences with respect to what 
is happening in the lower energies regime.
Tsang et al. \cite{TsangPRL92} have probed the isospin
diffusion mechanism for the systems $^{124}Sn +$ $^{112}Sn$ at
$E=50AMeV$ in a peripheral impact parameter range $b/b_{max}>0.8$,
observing the isoscaling features of the light
isotopes $Z=3-8$ emitted around the projectile rapidity. 
An incomplete equilibration has been deduced. The
value of the isoscaling parameter $\alpha=0.42 \pm 0.02$
for $^{124}Sn +$ $^{112}Sn$ differs substantialy from
 $\alpha=0.16 \pm 0.02$ for $^{112}Sn +$ $^{124}Sn$.
The isospin imbalance ratio \cite{RamiPRL84}, defined as

\begin{equation}\label{imb}
R_{i}(x) = \frac{ 2 x - {x}^{124+124} - {x}^{112+112}}
{{x}^{124+124} - {x}^{112+112}}
\end{equation}

($i=P,T$ refers to the projectile/target rapidity measurement,
and $x$ is an isospin dependent observable, here
the isoscaling $\alpha$ parameter) was estimated to be
around $R_P(\alpha)=0.5$ 
(vs.$R_P(\alpha)=0.0$ in full equilibration). 
This quantity can be sensitive to the
density dependence of symmetry energy term since the
isospin transfer takes place through the lower density
neck region.

\section{Neck Observables}

\subsubsection*{
Properties of Neck-Fragments, mid-rapidity $IMF$ produced in semicentral 
   collisions: correlations between $N/Z$, $alignement$ and size} 
 
The alignement between $PLF-IMF$ and $PLF-TLF$ directions
represents a very convincing evidence of the dynamical origin of the 
mid-rapidity fragments produced on short time scales \cite{BaranNPA730}. 
In fact a very selective correlation is provided by the anisotropy
in the fragment emission, measured by the in-plane azimuthal angle, 
$\Phi_{plane}$, defined as the angle between
the projection of the $PL-IMF$ scission axis onto the reaction plane and 
the separation axis between $PLF$ and $TLF$, \cite{StefaniniZFA351}. 
A $\vert \Phi_{plane} \vert \simeq 0$ collects events
corresponding to asymmetric ``fissions'' of the $PL$-system very aligned
along the outgoing $PL-TL$ separation axis. At variance a statistical 
fission dominance 
would correspond to a flat $\Phi_{plane}$ behavior.
The form of the
$\Phi_{plane}$ distributions (centroid and width) can give a direct
information on the fragmentation mechanism \cite{Dynfiss05}.

\subsubsection*{Time-scale measurements}

   The estimation of time scales for fragment formation from velocity
   correlations appears to be a very exciting possibility 
\cite{PaganoNPA734,Velcorr04}. With a good event by event detection of 
the $TL/PL$ residues we can measure the violations of the Viola systematics
for the $TLF-IMF$ and $PLF-IMF$ systems which tell us how much the $IMFs$
are uncorrelated to the $spectator$ remnants \cite{wilcz}.

\begin{figure}
\centering
\includegraphics*[scale=0.45]{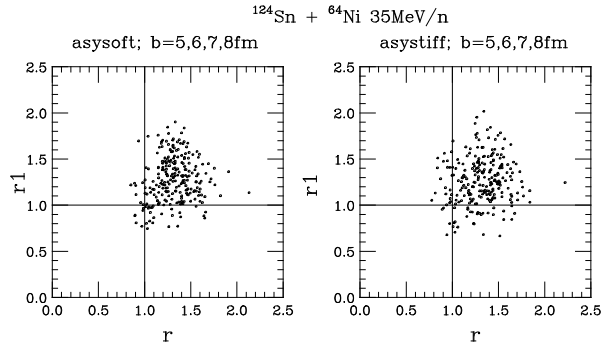}
\caption{Wilczynski-2 Plot: correlation between deviations from Viola 
systematics, see text. Results are shown for two $asy-EOS$.}
\label{wilcz2}
\end{figure}

For each Neck-$IMF$ we can evaluate the ratios
$r=v_{rel}(PLF)/v_{viola}(PLF)$, ($r1=v_{rel}(TLF)/v_{viola}(TLF)$).
In  Figure \ref{wilcz2}
we plot $r1$ against $r$ for each $NIMF$ for the $Sn+Ni$ reaction
 \cite{BaranNPA730}. 
We call such a representation
a $Wilczynski-2$ plot \cite{wilcz}. 
The solid lines represent the loci of the $PL$-($r=1$) and $TL$-($r1=1$)
 fission events respectively.
The values $(r,r1)$  appear simultaneously 
larger than $1$ suggesting a weak $NIMF$ correlation with {\it both} 
$PLF$ and $TLF$, in contrast to a statistical fission mechanism.
The process has some similarities
with the participant-spectator scenario. However the dynamics appear much 
richer than in the simple sudden abrasion model, where the locus of
the $r-r1$ correlation should be on mainly the bisectrix. Here the wide 
distributions
of Fig.\ref{wilcz2} reveal 
a broad range of fragment velocities, typical of the
instability evolution in the neck region
that will lead to large dynamical fluctuations on $NIMF$ properties. 

The $\Phi_{plane}$ distributions, corresponding to the same $NIMF$
events analysed before, are covering a quite
limited angular window, close to the full alignment configuration,
$\Phi_{plane}=0^{\circ}$. The distribution becomes wider when
we approach the two $r,r1~=~1$ lines of the Fig.\ref{wilcz2}.

With appropriate cuts in the velocity correlation plots 
and $\Phi_{plane}$ distributions
we
   can follow the properties of clusters produced from sources with
   a "controlled" different degree of equilibration. We can figure out
   a continous transition from fast produced fragments via neck instabilities
   to clusters formed in a dynamical fission of the Projectile(Target) 
   residues up to the evaporated ones (statistical fission). Along this 
   line it would be even possible to disentangle the effects of volume
   and shape instabilities. The isospin dynamics will look different 
   in the various scenarios and rather dependent on the symmetry
   term of the $EOS$.
 
\begin{figure}
\centering
\includegraphics*[scale=0.35]{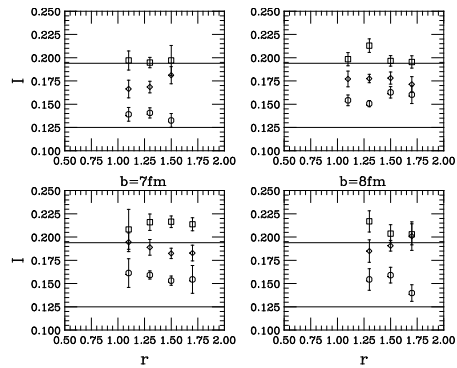}
\caption{$124Sn+64Ni$ at $35AMeV$. $NIMF$ isospin content for $asysoft$ 
(circles), $asystiff$ (rombs) 
and $superasystiff$ (squares)
$EOS$ as a function of $r$-deviation from Viola systematics, 
at impact parameters from $b=5fm$ to $8fm$. The two solid lines
represent the mean asymmetries of the projectile (top) and target (bottom). }
\label{papisorr}
\end{figure}

A more promising observable seems to be the isotopic content
of the Neck-$IMF$.
For the three $asy-EOSs$ introduced before we plot in 
Fig.\ref{papisorr} the average isotopic composition $I$ of
the $NIMF$'s as a function of the $PLF$ $r$-deviation from Viola
systematics. At all impact parameters clear differences 
are evident. The average asymmetry does not depend
strongly on $r$. We notice however that it increases 
with the stiffness of the symmetry potential around and below saturation. 
The {\it superasystiff}
parametrization, i.e. with an almost parabolic increasing behavior
around $\rho_0$, produces systematically more neutron rich
$NIMF$s.
This effect is clearly due to a different neutron/proton migration
at the interface between $PL/TL$ ''spectator'' zone of normal
density and the dilute neck region where the $NIMF$s are 
formed.
The Wilczynski-2 plot will
help to make the selections.  
We have to remind that all the results presented here refer to properties of
primary (excited) fragments. The neutron excess signal appears likely to be
washed out from later evaporation decays. A reconstruction of the
primary fragments with neutron coincidence measurements would be very
important.

\subsubsection*{Isospin Dynamics}

\begin{figure}[t]
\begin{center}
\includegraphics[bb = 16 277 563 552, scale=0.45]{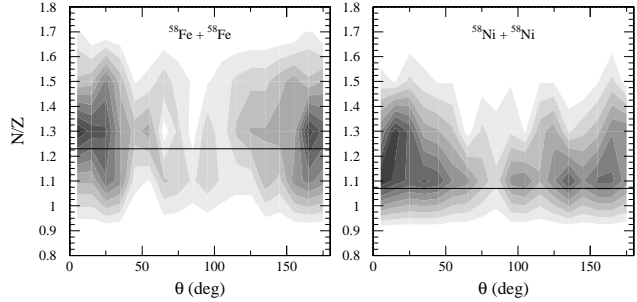}
\caption{$Fe,Fe$ vs. $Ni,Ni$ reactions at $47AMev$ ($b=4fm$): Contour 
plots of the N/Z distribution versus the emission angle.}
\label{NZ_angle}
\end{center}
\end{figure}

   Isospin effects on the reaction dynamics and $Isospin~Migration$:
   an interesting neutron enrichment of the overlap ("neck") region is
   expected, due to the neutron migration from higher (spectator) to 
   lower (neck) density regions. This effect is also nicely connected to 
   the slope of the symmetry energy. Neutron and/or light isobar measurements
   in different rapidity regions appear important.
   Moreover, moving from mid- to "spectator"- rapidities an increasing 
   hierarchy in the mass and $N/Z$ of the fragments is expected 
 \cite{LiontiarX05}.
   Some experimental evidences are in ref.\cite{ColinPRC67}. An interesting 
 related
   observable is the corresponding angular correlation due to the driving
   force of the Projectile(Target)-like partners \cite{LiontiarX05}.
The analysis of the $N/Z$ distribution versus the emission angle reveals
that larger fluctuations are present close to forward and backward angles,  
suggesting that IMF's that are more correlated to the spectator
matter may become more neutron-rich, see Fig.\ref{NZ_angle} for the
$Fe,Fe/Ni,Ni$ systems at $47AMeV$ \cite{LiontiarX05}.
 
We stress again that the $neck-IMF$ always present a neutron enrichment,
{\it even in the case of a n-poor system}. The latter paradox is due
to the pre-equilibrium isospin dynamics. 
In fact, as anticipated above, 
due to pre-equilibrium, the system will loose some protons 
and acquire a $N/Z$ larger than the initial one.  Then, before the di-nuclear
system reseparates, the neutron excess is transferred to the neck region
that is at low density.

\subsubsection*{Isospin Diffusion} 

Measure of charge equilibration in the "spectator" 
   region in semicentral collisions: $Imbalance~Ratios$ for different
   isospin properties. Test of the interplay between 
   concentration and density gradients in the isospin dynamics 
  \cite{BaranPR05,Isodiff05,ChenPRL05}.
   For the reasons noted before we expect to see a clear difference
   in the isospin diffusion between binary (deep-inelastic like)
   and neck-fragmentation events.

\begin{figure}
\begin{center}
\includegraphics*[scale=0.50]{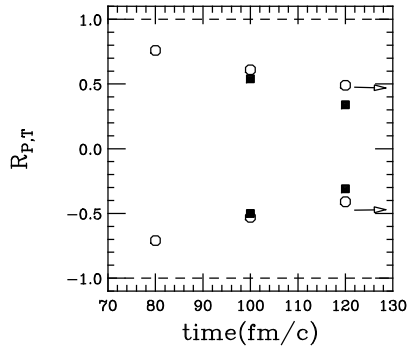}
\caption{$^{124}Sn+^{112}Sn$ $50~AMeV$ collision:
interaction time evolution of the projectile (upper) and target (lower)
isospin imbalance ratios. Squares: $Asysoft~ EOS$. Circles: $Superasystiff$.
The arrows correspond to the data of ref.\protect\cite{TsangPRL92}.}
\label{imbalance}
\end{center}
\end{figure}

We report its dependence on the interaction time (impact parameter)
for asysoft (squares)
and asysuperstiff (circles) $EOS$ in Fig.\ref{imbalance},
for the $Sn,Sn$ case \cite{BaranPR05,Isodiff05}.
A good degree of isospin equilibration corresponds to
a ``convergence'' to $0$ of both $R_{P}$ (upper curves) 
and $R_{T}$ (lower curves).
From our results we conclude that an asystiff-like
$EOS$ provides a better agreement to the experimental
observations, shown as arrows in Fig.\ref{imbalance}, 
 \cite{TsangPRL92}. 
In fact in the
$MSU$ experiment there is no particular selection on binary events.
We expect that the presence of events with  production of Neck-$IMFs$ 
will induce an apparent larger isospin equilibration since
more neutrons will migrate to the neck region (and viceversa for protons),
as discussed before.

A clear difference between the two equations of state
is evident especially for the longer interacting time, $b=8fm$ case. 
Smaller values
of the isospin imbalance ratios for asysoft $EOS$
point towards a faster equilibration rate.
In  ref. \cite{TsangPRL92} a possible explanation was
proposed related to the observation that below normal
density the asysoft $EOS$ has a larger value of symmetry
energy. Therefore an enhanced isospin equilibration
has to be triggered if the diffusion takes place at lower density.
In fact the mechanism of charge equilibration
is more complicated at these energies due to
reaction dynamics (fast particle emissions,
density gradients, etc.), with interesting compensation effects.

\subsubsection*{ ``Pre-equilibrium'' emissions}

As already remarked, the isospin content of the fast particle emission 
   can largely influence the subsequent reaction dynamics, in particular the
   isospin transport properties (charge equilibration, isospin diffusion).
   We can reach the paradox of a detection of isospin dynamics effects 
   in charge symmetric systems.

Finally the simultaneous measurements of properties of fast nucleon emissions
and of the neck dynamics can even shed lights on the very controversial
problem of the isospin momentum dependence \cite{BaranPR05,ChenPRL05}.

\subsubsection*{Outlook}
We stress the richness of the phenomenology and nice opportunities of
getting several cross-checks from completely different experiments.
Apart the interest of this new dissipative mechanism and the
amazing possibility of studying properties of fragments produced
on an almost continous range of time scales, we remark the
expected dependence on the isovector part of the nuclear $EOS$. 
From transport simulations we presently get some indications of 
"asy-stiff" behaviors, i.e. increasing repulsive density dependence of 
the symmetry term, but not more fundamental details. 
Moreover, all the available data are obtained with stable beams, i.e. 
within low asymmetries.

\subsection*{Acknowledgements}
This report is deeply related to ideas and results reached in
very pleasant and fruitful collaborations with extremely nice people:
V.Baran, M.Colonna, G.Fabbri, 
G.Ferini, Th.Gaitanos, V.Greco, R.Lionti, 
B.Liu, S.Maccarone, F.Matera, M.Zielinska-Pfabe', J.Rizzo, L.Scalone 
and H.H.Wolter. I have learnt a lot 
from all of them in physics as well as in human relationships.





\end{document}